\RequirePackage{fix-cm}
\documentclass[smallextended]{svjour3}       % onecolumn (second format)
\smartqed  % flush right qed marks, e.g. at end of proof

\usepackage [latin1]{inputenc}
\usepackage{graphicx}
\usepackage{amsmath}
\usepackage{marvosym}
\usepackage[noend]{algpseudocode}
\usepackage{algorithmicx,algorithm}
\usepackage{indentfirst}
\usepackage{color}
\usepackage{epstopdf}

\begin{document}

\title{ An Optimized Quantum Minimum Searching Algorithm with Sure-success Probability and Its Experiment Simulation with Cirq %\thanks{Grants or other notes
}
\subtitle{}


\author{Wenjie Liu\textsuperscript{1,2 \Letter}       \and
	    Qingshan Wu\textsuperscript{2}        \and
        Jiahao Shen\textsuperscript{2}        \and
        Jiaojiao Zhao\textsuperscript{2}      \and
        Mohammed Zidan\textsuperscript{3}    \and
        Lian Tong\textsuperscript{4}
 %etc.
}


\institute{   \Letter \quad wenjiel@163.com \\
              $^1 $ \quad  Engineering Research Center of Digital Forensics, Ministry of Education, Nanjing 210044, China \\
              $^2 $ \quad  School of Computer and Software, Nanjing University of Information Science and Technology, Nanjing 210044, China \\
              $^3 $ \quad  Center for Photonics and Smart Materials (CPSM), Zewail City of Science and Technology, October Gardens, 6th of October City, Giza 12578, Egypt\\
              $^4 $ \quad  School of Information Engineering, Jiangsu Maritime Institute, Nanjing 211100, China
}

\date{Received: date / Accepted: date}

\maketitle

\begin{abstract}
Finding a minimum is an essential part of mathematical models, and it plays an important role in some optimization problems. Durr and Hoyer proposed a quantum searching algorithm (DHA), with a certain probability of success, to achieve quadratic speed than classical ones. In this paper, we propose an optimized quantum minimum searching algorithm with sure-success probability, which utilizes Grover-Long searching to implement the optimal exact searching, and the dynamic strategy to reduce the iterations of our algorithm. Besides, we optimize the oracle circuit to reduce the number of gates by the simplified rules. The performance evaluation including the theoretical success rate and computational complexity shows that our algorithm has higher accuracy and efficiency than DHA algorithm. Finally, a simulation experiment based on Cirq is performed to verify its feasibility.

\end{abstract}

\section{Introduction}
\label{sec:1}
In recent years, the development of big data has made it urgent to deal with more data with higher speed and better efficiency. Therefore, some researchers have begun to try to use genetic algorithms(GA)\cite{Deb2002}, particle swarm optimization(PSO)\cite{Eberhart2001}, and some other state-of-art searching algorithms\cite{Ali2018,Tariq2019,Ali2020} to improve search efficiency. Using the properties of quantum mechanics, researchers have discovered some quantum algorithms to accelerate a series of algorithms in quantum computers \cite{Tao2019,Liu2019}. Besides, many researchers try to apply quantum mechanics in other fields, such as quantum key agreement(QKA)\cite{Huang2017,Liu2018Xu}, quantum secret sharing (QSS)\cite{Hillery1999,Liu2019Zhang}, blind quantum computation (BQC)\cite{Fitzsimons2017,Liu2019Xu}, quantum private query (QPQ) \cite{Gao2019,Liu2019GaoP}, and even quantum machine learning (QML)\cite{Schuld2019,Liu2020}.
Shor's factoring algorithm \cite{Shor1994} is a well-known example of a quantum algorithm outperforming the best known classical algorithm. This algorithm can effectively find discrete logarithms and factor integers on a quantum computer. In order to speed up the search problem, Grover proposed a quantum search algorithm \cite{Grover1996} in 1996. This algorithm can solve the searching problem by using approximately $\sqrt {N} $ operations rather than approximately $N$ operations in classical algorithm. Later, the database search algorithm gradually attracted wide attention of many scholars. In 2010, Diao pointed out that only if the ratio of the solution $ M $ to the database size $ N $ is 1/4, a strict and accurate search can be performed \cite{Diao2010}. Especially, the highest failure rate is 50\% when $M/N = 1/2$. To improve the efficiency of the Grover algorithm, researchers have explored various generalized and modified versions of the Grover algorithm,
including phase matching methods \cite{Long1999}, for an arbitrary initial amplitude distribution \cite{Biham1999}, recursion equations method \cite{Biham2000}, Grover-Long algorithm \cite{Long2001,Long2002} and fixed-point \cite{Grover2005}. Among them, Grover-Long algorithm has one adjustable phase that finds the target with zero failure rate for any database and with exactly the same number of iterations as Grover algorithm.

With the development of big data, finding a minimum or maximum is a significant issue in many fields. Classically, approximately $N $ operations are required for searching the maximum or minimum problems. But its quantum counterpart proposed by Durr and Hoyer \cite{Durr1996} achieves quadratic speedup, which was based on the quantum exponential searching algorithm \cite{WORONOWICZ2000,Boyer1998,Brassard1998,Castagnoli2016}. When the number of solutions is unknown, the quantum exponential searching algorithm reduces its failure rate at the expense of repeatedly performing Grover's algorithm with different number of iterations.However, Grover algorithm is not a sure-success algorithm. Besides, the operation of marking state in the repetition approach also takes time. To solve these problems, we propose an optimized quantum  minimum searching algorithm (OQMSA). The main contributions are as follows:

1. We utilize Grover-Long searching to implement the optimal exact searching, and then propose a sure-success quantum minimum searching algorithm.

2. In order to improve the efficiency of our algorithm, a dynamic strategy is proposed to reduce the iterations. The ratio of the solutions to the size of dataset is different, then we use different minimum searching methods.

3. In terms of quantum circuit implementation, we propose two simplified rules for the oracle operation, thereby reducing the number of quantum gates.

The remainder of this paper is organized as follows. In Section 2, we review DHA algorithm. In Section 3, we present OQMSA based on Grover-Long algorithm and the general quantum circuits of key steps. In Section 4, we analyze the success rate and complexity of OQMSA. In Section 5, an experiment based on Cirq framework to solve a specific problem is presented, which shows that OQMSA is indeed more efficient. In Section 6, we draw a conclusion and look forward to its future applications.

\section{Review of DHA Algorithm}
\label{sec:2}
DHA algorithm \cite{Durr1996} can solve the problem of finding the minimum of unsorted database. It aims to find the index of a smaller item than the value determined by a particular threshold index. Then the result is chosen as the new threshold. This process repeats for times to increase the probability of finding the index of the minimum. If there are $t \ge 1$ marked table entries, the quantum exponential searching algorithm will return one of them with equal probability after an expected number of  $O(\sqrt {N/t} )$ iterations. If no entry is marked, it will run forever. The steps of DHA algorithm are as follows:
\\ \hspace*{\fill} \\
Step 1: Choose threshold index $0 \le y \le N - 1$ uniformly at random.
\\ \hspace*{\fill} \\
Step 2: Repeat the following and interrupt it when the total running time is more than  ${\rm{22}}{\rm{.5}}\sqrt N  + 1.4{\lg ^2}N $.

(a) Prepare the initial state $\sum\nolimits_j {\frac{1}{{\sqrt N }}\left| j \right\rangle \left| y \right\rangle }$. Mark every item $j $ when $D[j] <   D[y]$ .

(b) Apply the quantum exponential searching algorithm on the initial state.

(c) Measure the first register: output $y' $ if  $D[y'] <  D[y]$ , then set threshold index $y $  to $y'$ .
\\ \hspace*{\fill} \\
Step 3: Return the measure value $y$.

DHA algorithm provides a simple quantum algorithm which solves the problem using  $O\sqrt N $ probes.
The main subroutine is the quantum exponential searching algorithm, which is a generalization of Grover algorithm.
However, we find the uncontrollability of deflection angle in origin Grover algorithm, which result in the low success rate in DHA algroithm. In the meantime, the construction of the oracle is so complicated that the complexity raises a lot. To improve the success rate and decrease the complexity, we propose an optimized quantum minimum searching algorithm.

\section{An optimized quantum minimum searching algorithm with sure-success probability }
\label{sec:3}

\subsection{The proposed OQMSA algorithm}
\label{sec:3.1}
%stepÈ¥µô
In order to conduct the exact searching i.e., obtain the minimum with a success rate close to 1, we propose an optimized quantum minimum searching algorithm. Suppose that $D$ is an unsorted database with N items, $D = \{ {d_j}|0 \le j < N{\rm{\} }}$. All data items all are encoded into a n-qubit superposition state $\left| \varphi  \right\rangle $. Algorithm \ref{alg1} gives the specific steps of our OQMSA algorithm.

\begin{algorithm}[t]

\caption{The Proposed OQMSA Algorithm for finding the minimum}
\label{alg1}
\hspace*{0.02in} {\bf Input:}
Database $D = \{ {d_j}|0 \le j < N\} $, a random item $d' \in D$.

\hspace*{0.02in} {\bf Output:} %
The minimum item ${d_{min }}$.

\begin{algorithmic}[1]
\For{$i = 0$ \textbf{to} $\left\lceil {LogN} \right\rceil $}
    \State ${t_{max}} = [(\pi /2 - \arcsin (1/\sqrt N ))/\arcsin (1/\sqrt N )]$ ;
    \State $t = 1$, $\lambda  = \frac{6}{5}$, and $r \to  + \infty $;
        \While{($t \le {t_{max}}$ \&\& $r > d'$)}
        \State Prepare the initial state $ \left| \varphi  \right\rangle {\rm{ = }}\frac{1}{{\sqrt N }}\sum\limits_{j = 0}^{N - 1} {\left| {{d_j}} \right\rangle } $  according to $N$ data items;
¡¡¡¡          \If {$\frac{{M}}{N} > \frac{1}{9}$}
                    \State $t' = randint(0,{\rm{ }}\left\lceil t \right\rceil )$;
                    \State Apply Grover-Long searching on $\left| \varphi  \right\rangle $ with $t'$ iterations $ \to \left| {\varphi '} \right\rangle $;
                    \State $t = t*\lambda $;
              \Else
                    \State Apply Grover-Long searching on $\left| \varphi  \right\rangle $  with ${t_{max}}$ iterations $ \to \left| {\varphi '} \right\rangle $;
¡¡¡¡          \EndIf
                \State \textbf{end}
              \State Measure $\left| {\varphi '} \right\rangle $ and assign the measurement result to $r$;
        \EndWhile¡¡
        \State \textbf{end}
    \If{$r < d'$}
¡¡¡¡¡¡¡¡       \State $d'=r $;
                \State $i=0$;
¡¡¡¡\EndIf
    \State \textbf{end}
\EndFor
\State \textbf{end}
\State \Return ${d_{min }} = d'$.
\end{algorithmic}
\end{algorithm}

\begin{algorithm}[t]
\caption{Grover-Long searching} % Ëã·¨µÄÃû×Ö
\hspace*{0.02in} {\bf Input:}
initial state $\left| \varphi  \right\rangle $, $t$, $d'$.

\hspace*{0.02in} {\bf Output:} %Ëã·¨µÄ½á¹ûÊä³öe
finial state $\left| {{\varphi'}} \right\rangle $.
\begin{algorithmic}[1]
    \State $\phi  = 2\arcsin (\frac{{\sin \frac{\pi }{{4t + 2}}}}{{\sqrt {M/N} }})$;
    \State $\left| {{\varphi _0}} \right\rangle  = \left| \varphi  \right\rangle $;
    \For{$k = 0$ \textbf{to} $t$}
    \State Apply the oracle operation on ${\rm{     }}\left| {{\varphi _k}} \right\rangle  \to \left| {\tilde \varphi } \right\rangle  = {e^{i\phi }}\sum\limits_{{d_m} \le d'}^{} {\left| {{d_m}} \right\rangle }  + \sum\limits_{{d_n} > d'}^{} {\left| {{d_n}} \right\rangle }$
    \State Apply the phase reverse operation on
    $\left| {\tilde \varphi } \right\rangle  \to \left| {{\varphi _{k + 1}}} \right\rangle $
    \EndFor
    \State \textbf{end}
    \State Return ${\rm{ }}\left| {\varphi'} \right\rangle  = \left| {{\varphi _{k + 1}}} \right\rangle $.
\end{algorithmic}
\end{algorithm}

As shown in \textbf{Algorithm 1}, we firstly prepare the initial state $\left| \varphi  \right\rangle  = \frac{1}{{\sqrt N }}\sum\limits_{j = 0}^{N - 1} {\left| {{d_j}} \right\rangle } $ according to $N$ data items. Then, the dynamic strategy is used to increase the probability of target state. The dynamic strategy is as follows: When $M/N > 1/9  $, the number of iterations $t'$ is an integer chosen randomly range from 0 to $\left\lceil t \right\rceil $ ($\left\lceil t \right\rceil $ is max number of current iteration). And then, the $t'-$iteration Grover-Long searching (as shown in \textbf{Algorithm} 2) is applied on $\left| \varphi  \right\rangle$. If $M/N < 1/9  $, the $t_{max}-$iteration Grover-Long searching is applied on $\left| \varphi  \right\rangle$. Finally, we obtain the measure value $r$ and compare $r$ with $d'$, the $d'$ will be replaced if $r<d'$. This process will repeat until we get same current minimum for $\left\lceil {LogN} \right\rceil $ times, and the algorithm will end with returning $d_{min}$.

\subsection{Quantum circuit and its optimization}
\label{sec:3.2}
\paragraph{A. Oracle operation}
~\\
\indent The oracle operation plays a role in identifying and marking the target state in the circuit. The oracle's marker factor is a phase rotation that changes the amplitude in front of the target state to ${e^{i\phi }}$ . The oracle can be described as diagonal matrix that only has ${e^{i\phi }}$ and
$1$, as shown in Eq.(\ref{equ:6})
\begin{equation}
O = {e^{i\phi }}\sum {\left| \upsilon  \right\rangle \langle \upsilon |}  + \sum\limits_{\tau  \ne \upsilon } {\left| \tau  \right\rangle } \langle \tau | ,
\label{equ:6}
\end{equation}
where $\nu $ is the state which need to be marked.

\noindent \textbf{Rule 1.} When $d' = {2^m} - 1$, where $d'$ is current minimum, the oracle can be simplified as below:
\begin{equation}
\begin{split}
O(m) =& ({e^{i\phi }}\left| {{0_0}} \right\rangle \left\langle {{0_0}} \right|+{\left| {{1 _0}} \right\rangle \left\langle {{1 _0}} \right|} ) \otimes {I^{ \otimes m}}.
\end{split}
\end{equation}

\begin{figure}
\centering
\includegraphics[width=1.0\linewidth]{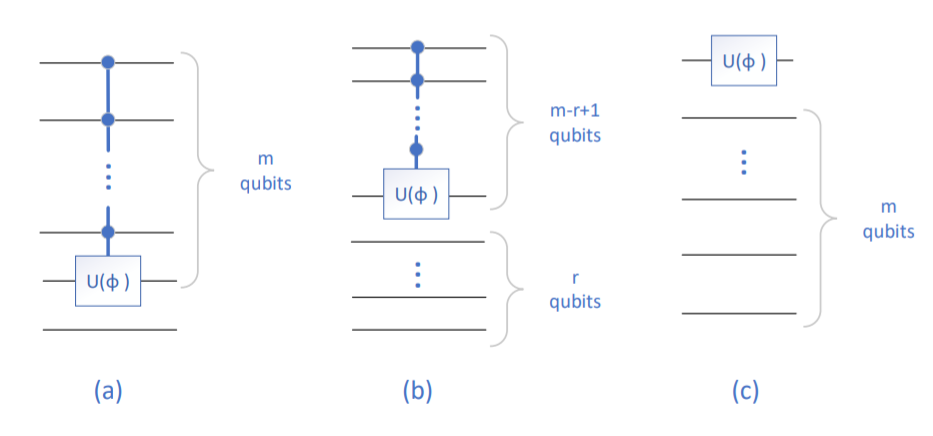}
\caption{The schematic diagram of the first simplified rule. (a) The circuit for marking two continuous states. (b) The circuit for marking ${2^r}$ continuous states. (c) The circuit for marking $2^m$ continuous states.}
\label{fig:1}
\end{figure}

Since $d'={2^m} - 1$, we encode $d'+1$ into quantum state $\left| {{1_0}} \right\rangle  \otimes \left| {{0_1}{0_2} \cdots {0_m}} \right\rangle $. The first qubit of all solutions must be $\left| 0 \right\rangle $, they can be marked only if we ensure that the first qubit is $\left| 0 \right\rangle $. The oracle will become an operation that only change the amplitude of state where the first qubit is $\left| 0 \right\rangle $ to ${e^{i\phi }}$. The schematic diagram is shown in Fig.~\ref{fig:1}.

\noindent \textbf{Rule 2.} When $d' \ne {2^m} - 1$, $d' + 1 = \sum\limits_{i = 0}^{n-1} {{a_i} \cdot {2^{n-i-1}}} $, ${a_i} \in \{ 0,1\}$, the oracle should be simplified iteratively through Algorithm 3.

\begin{algorithm}[t]
\caption{The iterative simplified algorithm}
\hspace*{0.02in} {\bf Input:}
The binary string of $d',({a_0}{a_1} \cdots {a_{n - 1}})$.

\hspace*{0.02in} {\bf Output:} %Ëã·¨µÄ½á¹ûÊä³öe
$ O $.
\begin{algorithmic}[1]
    \State $i=0$;
    \While {($i < n$)}
        \If {${a_i} = 1$}
            \If {$i=0$}
¡¡¡¡            \State Obtain $O'(i)=O(n-i-1)$ referring to \textbf{Rule 1};
            \Else
                \State Obtain $O'(i)=\left| {{a_0}{a_1} \cdots {a_{i - 1}}} \right\rangle \left\langle {{a_0}{a_1} \cdots {a_{i - 1}}} \right| \otimes O(n - i - 1)$;
            \EndIf
        \Else
            \If {$i=0$}
¡¡¡¡            \State Obtain $O'(i)=\left| 1 \right\rangle \left\langle 1 \right| \otimes {I^{ \otimes n-i-1}}$;
            \Else
                \State Obtain $O'(i)=\left| {{a_0}{a_1} \cdots {a_{i - 1}}} \right\rangle \left\langle {{a_0}{a_1} \cdots {a_{i - 1}}} \right| \otimes \left| 1 \right\rangle \left\langle 1 \right| \otimes {I^{ \otimes n-i-1}}$;
            \EndIf
¡¡¡¡    \EndIf
        \State $i=i+1$;
    \EndWhile
\State Obtain $O'(i) = \left| {{a_0}{a_1} \cdots {a_{n - 1}}} \right\rangle \left\langle {{a_0}{a_1} \cdots {a_{n - 1}}} \right|$
\State return $O=\sum {O'(i)} $
\end{algorithmic}
\end{algorithm}

\begin{figure}
\centering
\includegraphics[width=1.0\linewidth]{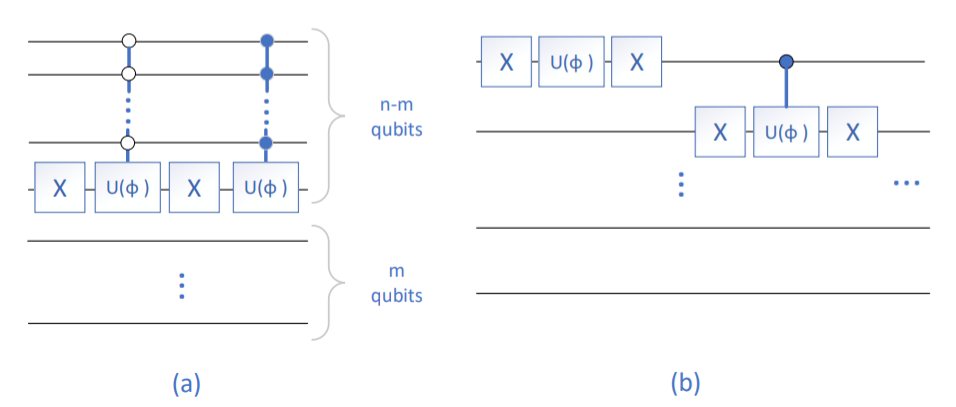}
\caption{A schematic diagram of the second equivalent simplified rule. (a) The original circuit. (b) The simplified circuit.}
\label{fig:2}
\end{figure}

\begin{figure}
\includegraphics[width=1\linewidth]{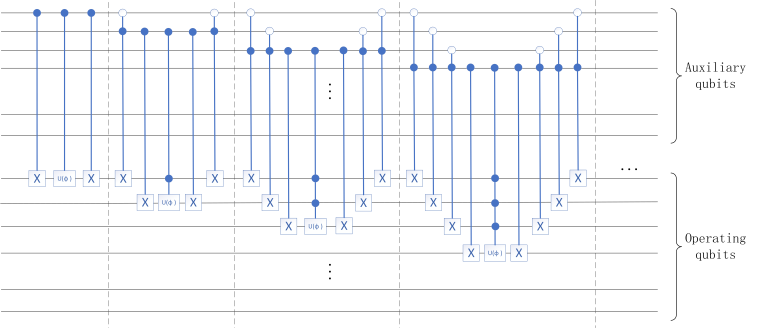}
\caption{The optimized circuit for the oracle in OQMSA.}
\label{fig:3}
\end{figure}

For the case $d' \ne {2^m} - 1$, $d' + 1 = \sum\limits_{i = 0}^{n-1} {{a_i} \cdot {2^{n-i-1}}} $, $d' + 1$ is encoded into $\left| {{a_1}{a_2} \cdots {a_{n - 1}}} \right\rangle $. Algorithm 3 will determine which states need to be marked by iterating on a binary string of ${{a_1}{a_2} \cdots {a_{n - 1}}}$. Then the corresponding quantum circuit can be divided into many sub-circuits such as Fig.~\ref{fig:2}(a). Finally, we can obtain the oracle which is used to mark all solutions after n iterations and it can be simplified as Fig.~\ref{fig:2}(b). The general optimized oracle circuit is showed in the Fig.~\ref{fig:3}.

For instance, when $d' = 31 = 2^5-1$ and n=6, we simplify the oracle as $O(5) = ({e^{i\phi }}\left| 0 \right\rangle \left\langle 0 \right| + \left| 1 \right\rangle \left\langle 1 \right|) \otimes {I^{ \otimes 5}}$ according to Rule 1. We only need to operate on the first qubit to mark all the items range from 0(000000) to 31(011111). When $d' = 35 = 2^5 + 2^2 - 1$, we can obtain the oracle($({e^{i\phi }}\left| 0 \right\rangle \left\langle 0 \right| + \left| 1 \right\rangle \left\langle 1 \right|) \otimes {I^{ \otimes 5}} + \left| {100} \right\rangle \left\langle {100} \right| \otimes ({e^{i\phi }}\left| 0 \right\rangle \left\langle 0 \right| + \left| 1 \right\rangle \left\langle 1 \right|) \otimes {I^{ \otimes 2}}$) according to \textbf{Rule} 2. Where the oracle mark the items range from 0(000000) to 31(011111), and then mark the items between 32(100000) and 35(100011). In this way, all states $\le$ 35 are marked out.

\begin{figure}
\includegraphics[width=0.5\textwidth]{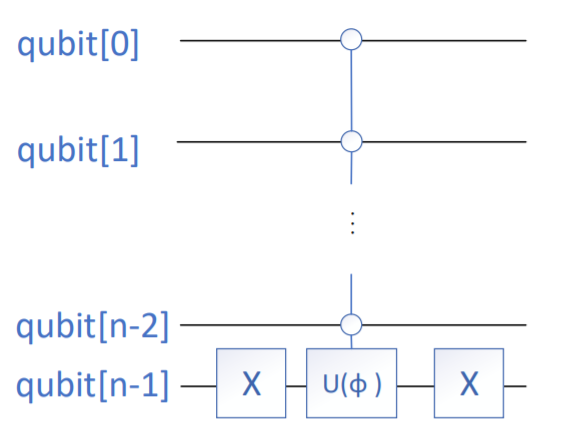}
\caption{The general circuit for $I_{0}$ operation.}
\label{fig:4}
\end{figure}

\paragraph{B. Phase deflection}
~\\
\indent The operation of phase deflection use ${e^{i\phi }}$ to distinguish the states, because the oracle use ${e^{i\phi }}$  to mark the target states. Besides, this operation can be divided into three parts: ${W^{ - 1}}$, ${I_0}$ and $W$, where $W$ is an operation of preparing initial state, ${I_0}$ is the core of the whole phase deflection. The operator of $I_{0}$ can be described as a diagonal matrix, as shown in Eq. \ref{equ:3}.
\begin{equation}
{I_0} = {e^{i\varphi }}|0\rangle \left\langle 0 \right| + \sum\nolimits_{\tau  = 1}^{{2^{n{\rm{ - }}1}}} {\left| \tau  \right\rangle \langle \tau |}  = {\mathop{\rm diag}\nolimits} \left[ {{e^{i\varphi }},1, \ldots ,1} \right]_{{2^n}}^{} ,
\label{equ:3}
\end{equation}
where $n$ is the number of qubits.
$I_{0}$ can be converted to the quantum circuit, as shown in Fig. \ref{fig:4}.

\section{Performance Evaluation}
\label{sec:4}
We analyze the performance of OQMSA from two aspects: the success rate and the computation complexity.

\subsection{Success Rate}
\label{sec:4.1}
As described in the literature \cite{Durr1996}, the success rate of the DHA algorithm is slightly greater than 0.5. Through the following derive, we can know that our algorithm has a higher success rate. First of all, if we want to obtain the exact value of ${t_{max }}$, $\varphi$, $\beta$, we must know $M /N$. However, the dataset is not directly available. When a random value $d',$ is selected, we do not know how many solutions. We replace the exact values $M, N$ with estimated values $\widetilde{M}, \widetilde{N}$. Therefore, the unknown database has $\widetilde{N}=2^{n}$ non-repeating values and the estimated number of marked states is $\widetilde{M}=$ $\mathrm{d'}+1 .$ The distribution function of the estimated database is regarded as a uniform distribution, where the probability density function is $\tilde{\rho}(x)=\frac{1}{\sqrt{N}} .$ And the distribution function of the real database is unknown. It is important to quantify the impact of the gap between $\frac{\widetilde{M}}{\widetilde{N}}$ and $\frac{M}{N}$ on the failure rate $\varepsilon_{G L}$.

The cumulative distribution function of the estimated database in $\left[0, \mathrm{d'}\right]$ is described as:

\begin{equation}
\widetilde{P}(x)=\int_{0}^{d_{0}} \widetilde{p}(x) d x=\frac{\widetilde{M}}{\widetilde{N}} .
\end{equation}
Similarly, the cumulative distribution function of the real database $P(x)$ is $\frac{M}{N} .$ Therefore, if the actual database follows the uniform distribution in  $\left[0,2^{n}\right],$ then $\frac{\widetilde{M}}{\widetilde{N}}: \frac{M}{N} \approx 1 .$ In other words, the success rate of our algorithm is close to 1.

\subsection{Complexity}
\label{sec:4.2}
Note that the complexity of OQMSA is primarily composed of a total number of Grover-Long iterations and
the initial state preparation. So we calculate the complexity without other steps.
One main loop possesses $ t_{max} $ Grover-Long iterations. $ t_{max} $ can be described as Eq. \ref{eq:6.1}
\begin{equation}
t_{max} = floor\left( {\frac{{\frac{\pi }{{\rm{2}}}{\rm{ - }}\arcsin \left( {\sqrt {\frac{M}{N}} } \right)}}{{\arcsin \left( {\sqrt {\frac{M}{N}} } \right)}}} \right) + 1 .
\label{eq:6.1}
\end{equation}

For convenience, we consider the case of an infinity database, so
\begin{equation}
\mathop {\lim }\limits_{\sqrt {M/N}  \to 0} \arcsin \left( {\sqrt {\frac{M}{N}} } \right) \approx \sqrt {\frac{M}{N}} .
\label{eq:6.2}
\end{equation}

We can simplify the complexity of Grover-Long algorithm as Eq. \ref{eq:6.3}.
\begin{equation}
t_{max} = (\frac{\pi }{2} - 1{\rm{ + }}1) \times \sqrt {\frac{N}{M}}  = \frac{\pi }{2}\sqrt {\frac{N}{M}} .
\label{eq:6.3}
\end{equation}

The total of Grover-Long iterations can be described as Eq. \ref{eq:6.4}.
\begin{equation}
{R_G} = \sum\limits_{k = 0}^{K - 1} {{J_k} = } \frac{\pi }{2}\sum\limits_{k = 0}^{K - 1} {\sqrt {\frac{N}{{{M_k}}}} }   .
\label{eq:6.4}
\end{equation}
where $K $  represents the total number of main loops, ${M_k}$ represents the number of marked states of the $k$-th main loop. Since the marked states have the same amplitude after applying Grover-Long algorithm, the number of ($k+1$)-th main loop's marked quantum states is nearly half of the ($k$)-th main loop's. Thus, the complexity of all Grover-Long algorithm iterations can be described as Eq. (\ref{eq:6.5}):
\begin{equation}
\begin{array}{l}
{R_G} = \frac{\pi }{2}\left( {\sqrt {\frac{N}{{{M_0}}}}  + \sqrt {\frac{{2N}}{{{M_0}}}}  + ... + \sqrt {\frac{{{M_0}N}}{{{M_0}}}} } \right) = \frac{\pi }{2}\frac{{\sqrt {\frac{N}{{{M_0}}}}  \times \left( {1 - \sqrt {2{M_0}} } \right)}}{{1 - \sqrt 2 }}\\
 = \frac{\pi }{2}\left( {\sqrt 2  + 1} \right)\left( {\sqrt {2N}  - \sqrt {\frac{N}{{{M_0}}}} } \right) .
\label{eq:6.5}
\end{array}
\end{equation}

Next, we consider the complexity of the initial state preparation. Since it needs to be executed about ${\log _2}N$ times and each execution takes ${\log _2}N$ steps, it can be described as Eq.(\ref{eq:6.6})
\begin{equation}
{R_{init}} = {({\log _2}N)^2} .
\label{eq:6.6}
\end{equation}

We can calculate the complexity as Eq.(\ref{eq:6.7})
\begin{equation}
R = {R_G} + {R_{init}} = \frac{\pi }{2}\left( {\sqrt 2  + 1} \right)\left( {\sqrt {2N}  - \sqrt {\frac{N}{{{M_0}}}} } \right) + {({\log _2}N)^2} .
\label{eq:6.7}
\end{equation}
As claimed in Ref. \cite{Durr1996}, the complexity of DHA is ${\rm{22}}{\rm{.5}}\sqrt N  + 1.4{(\log_2 N)^2}$, while the complexity of our algorithm is $\frac{\pi }{2}(\sqrt 2  + 1)(\sqrt {2N}  - \sqrt {\frac{N}{{{M_0}}}} ) + {({\log _2}N)^2}$. Because $\frac{\pi }{2}(\sqrt 2  + 1)(\sqrt {2N}  - \sqrt {\frac{N}{{{M_0}}}} ) + {({\log _2}N)^2} < 2*3*2\sqrt N  + {({\log _2}N)^2} < 22.5\sqrt N  + 1.4{({\log _2}N)^2}$, our algorithm has a smaller time complexity.

\begin{figure}
\includegraphics[width=1\textwidth]{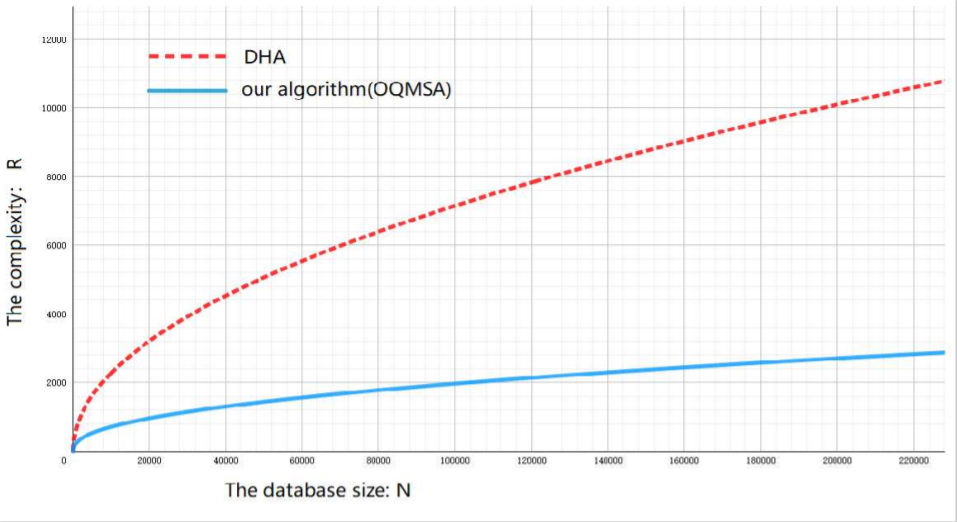}
\caption{Complexity comparison between DHA and our algorithms.}
\label{fig:5}
\end{figure}

Finally, we compare DHA algorithm with our algorithm under the same conditions. For convenience, we assume that ${M_0} = \frac{1}{2}N$.The complexity comparison of the two algorithms is shown in the Fig. \ref{fig:5}. It can be seen from the figure that as the algorithm increases, our algorithm has a greater advantage in the complexity of algorithm. Because OQMSA requires fewer gates in Oracle and decreases the number of iteration in dynamic construction of circuit with two simplified rules.

\section{Experiment Simulation }
\label{sec:5}
Cirq\cite{Cirq}, launched by Google in 2018, is a software library for writing, manipulating, and optimizing quantum circuits and then running them against quantum computers and simulators. Cirq attempts to expose the details of hardware, instead of abstracting them away, because, in the Noisy Intermediate-Scale Quantum (NISQ) regime, these details determine whether or not it is possible to execute a circuit at all. In order to verify the feasibility of our algorithm, we choose cirq as the experiment simulation platform, and design and implement a 6-qubits simulation experiment to perform the minimum searching in a data set (such as Table~\ref{table:1} or Table~\ref{table:2} ). Our experiments are conducted on a computer equipped with Intel Xeon 5218, double 2.3Ghz CPU, 64G RAM, and the version of Cirq is 5.0 under python 3.6.5.
\begin{table}
\caption{Dataset A for searching the minimum }
\label{table:1}       % Give a unique label
\begin{tabular}{llllll}
\hline\noalign{\smallskip}
Value & After Encoding & Value & After Encoding & Value & After Encoding \\
\noalign{\smallskip}\hline\noalign{\smallskip}
2     & 000010         & 24    & 011000         & 44    & 101100         \\
18    & 010010         & 40    & 101000         & 60    & 111100         \\
34    & 100010         & 56    & 111000         & 14    & 001110         \\
50    & 110010         & 10    & 001010         & 30    & 011110         \\
6     & 000110         & 26    & 011010         & 46    & 101110         \\
22    & 010110         & 42    & 101010         & 62    & 111110         \\
38    & 100110         & 58    & 111010         & 3     & 000011         \\
54    & 110110         & 12    & 001100         & 19    & 010011         \\
8     & 001000         & 28    & 011100         & 35    & 100011         \\
51    & 110011         & 7     & 000111         & 23    & 010111         \\
9     & 001001         & 55    & 110111         & 39    & 100111         \\
25    & 011001         & 41    & 101001         & 57    & 111001         \\
43    & 101011         & 27    & 011011         & 11    & 001011         \\
59    & 111011         & 13    & 001101         & 29    & 011101         \\
15    & 001111         & 61    & 111101         & 45    & 101101         \\
31    & 011111         & 47    & 101111         & 63    & 111111         \\
\noalign{\smallskip}\hline\noalign{\smallskip}
\end{tabular}
\end{table}

\begin{table}
\caption{Dataset B for searching the minimum }
\label{table:2}       % Give a unique label
\begin{tabular}{llllll}
	\hline\noalign{\smallskip}
	Value & After Encoding & Value & After Encoding & Value & After Encoding \\
	\noalign{\smallskip}\hline\noalign{\smallskip}
	45    & 101101      & 46    & 101110         & 42    & 101010         \\
	37    & 100101      & 38    & 100110         & 34    & 100010         \\
	21    & 010101      & 22    & 010110         & 18    & 010010         \\
	61    & 111101      & 62    & 111110         & 58    & 111010         \\
	53    & 110101      & 54    & 110110         & 50    & 110010         \\
	5     & 000101      & 6     & 000110         & 2     & 000010         \\
	44    & 101100      & 40    & 101000         & 47    & 101111         \\
	36    & 100100      & 32    & 100000         & 39    & 100111         \\
	20    & 010100      & 16    & 010000         & 23    & 010111         \\
	60    & 111100      & 56    & 111000         & 63    & 111111         \\
	52    & 110100      & 48    & 110000         & 55    & 110111         \\
	4     & 000100      & 0     & 000000         & 7     & 000111         \\
	\noalign{\smallskip}\hline\noalign{\smallskip}
\end{tabular}
\end{table}

\begin{figure}
	\includegraphics[width=1.0\linewidth]{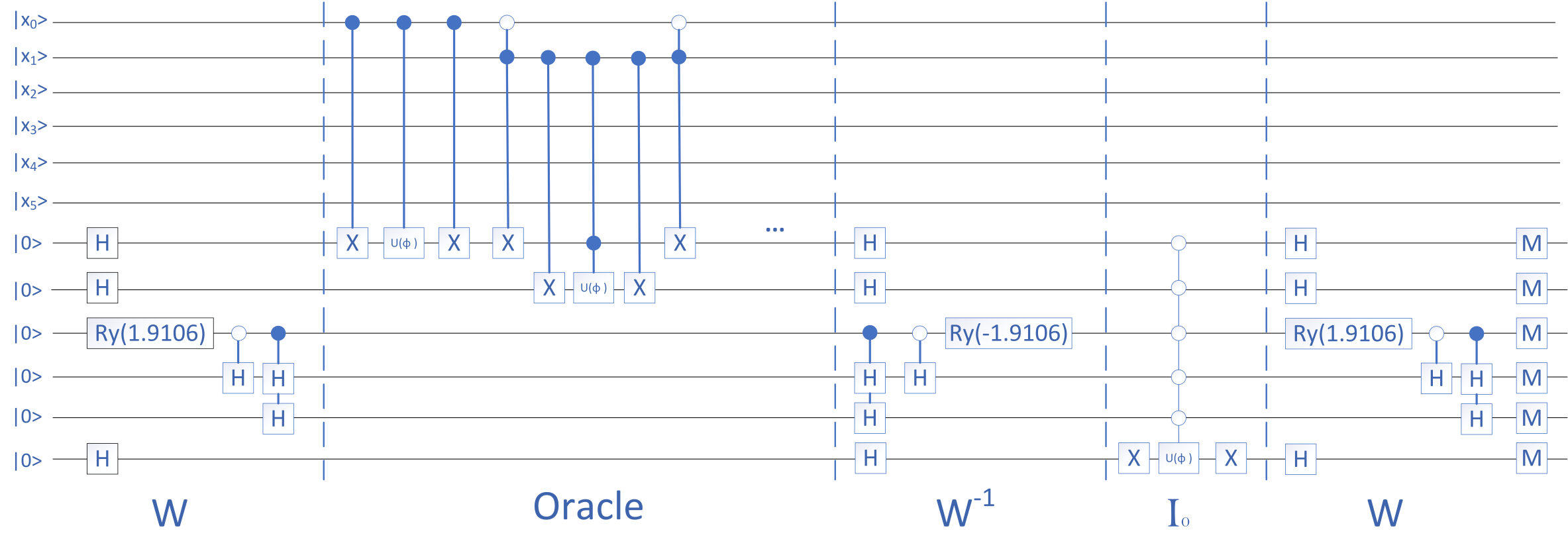}
	\caption{A 12-qubits circuit of Grover-Long algorithm with 48 items.}
	\label{fig:6}
\end{figure}

In beginning of simulation, a random item is selected to be the current minimum $d'$, the unsorted database is encoded into $n$ qubits according binary encoding method, $d'$ is also encoded into auxiliary qubit $\left| {{x_0}{x_1}{x_2}{x_3}{x_4}{x_5}} \right\rangle$, the amplitude of each state is the same value $\frac{{\rm{1}}}{{\sqrt N }}$. In the early stage of the algorithm, because the current minimum is likely to be very large, we choose the program of increasing the number of iterations dynamically to detect the upper bound in one algorithm. The advantage of this scheme is that we can get more accurate phase deflection without knowing how many solution sets there are. When the current minimum value decreases gradually, the previous scheme will cause the problem of building circuits many times. In this case, we set a threshold $1/9$. When ${d'}/N$ is less than $1/9$, the advantages of the original algorithm are incarnated. We estimated the number of solutions $ {\widetilde{M}} = {d'} + 1$ and the estimated database size $\widetilde{N} = {2^n}$, where $n=6$. At the end of an algorithm, we can obtain a measurement result $r$. If $r \le {d'}$, we replace the current minimum with ${r}$ and the algorithm is thought to operate successfully. By constantly changing the current minimum value, we will finally get the minimum value with a great probability. The complete circuit is shown in Fig. \ref{fig:6}.

\begin{figure}
	\includegraphics[width=0.85\linewidth]{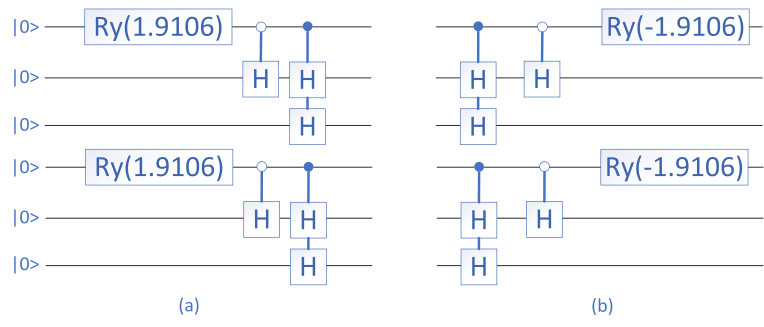}
	\caption{$W$ and  ${W^{-1}}$ operation of Dataset B. (a) The $W$ operation of Dataset B. (b) The ${W^{-1}}$ operation of Dataset B.}
	\label{fig:7}
\end{figure}

\begin{table}
	\caption{The measurement results of 20 experiments}
	\label{table:3}       % Give a unique label
	\begin{tabular}{lllll}
		\hline\noalign{\smallskip}
		      &     Dataset A &   &     Dataset B & \\
		\noalign{\smallskip}\hline\noalign{\smallskip}
		sequence   & minimum	 & others & minimum & others \\
		\noalign{\smallskip}\hline\noalign{\smallskip}
		1 & 984 & 16 & 985 & 15 \\
		2 & 987 & 13 & 979 & 21 \\
		3 & 986 & 14 & 980 & 20 \\
		4 & 982 & 18 & 986 & 14 \\
		5 & 984 & 16 & 976 & 24 \\
		6 & 984 & 16 & 982 & 18 \\
 	    7 & 988 & 12 & 985 & 15 \\
	    8 & 985 & 15 & 980 & 20 \\
		9 & 987 & 13 & 978 & 22 \\
	   10 & 986 & 14 & 983 & 17 \\										
		\noalign{\smallskip}\hline
	\end{tabular}
\end{table}

\begin{figure}
\centering
\includegraphics[width=1.0\linewidth]{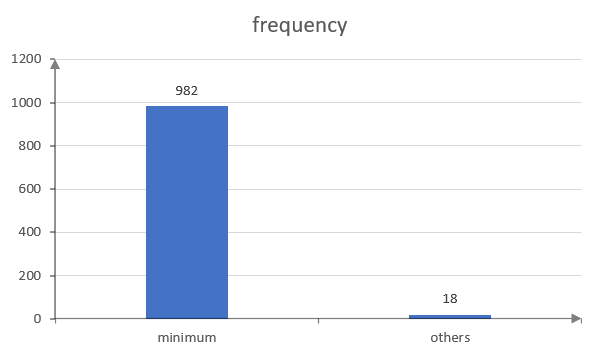}
\caption{The average consequence of 20 experiments.}
\label{fig:8}
\end{figure}

When the circuit is executed 1000 times, the average experimental results are shown in Fig. \ref{fig:8}, the minimum represents the result of finding the minimum successfully and the others deputies the failure results of the experiment. As shown in the chart, we get the minimum for 982 times in one thousand experiments. Obviously, this consequence proves that our algorithm has a high success rate. Even if we can't find the minimum with 100 percent probability, we still think that the algorithm is efficient under some conditions.The results of 20 experiments are shown in Table~\ref{table:3}.

\section{Conclusion}
\label{sec:6}
Based on Grover-Long algorithm, we proposed OQMSA, which is an improved version of DHA algorithm. Compared with classical algorithm, we have show the advantages of quantum algorithms in finding the minimum values to alleviate some of the challenges brought by the rapidly increasing amount of data. Besided, as the size of the database increases, it has a higher probability of success and a greater advantage in terms of complexity than DHA algorithm. In addition, we provide corresponding general-purpose quantum circuits. The optimized circuits are easy to implement on any general-purpose quantum computer. Meanwhile, the general circuit design methods can be implemented on the quantum platform. We demonstrate the advantage of our algorithm through a group 12-qubit experiment(6 of them are auxiliary bits) which is executed on Cirq platform and a real issue that is numerical simulated. In addition to the computational tasks we show in this paper, the algorithm can be a subroutine in other quantum algorithms that need to find a minimum value.

We hope that the theoretical and experimental results we present here well push further research and motivate innovations of other mathematical models. The paradigm combing classical steps and quantum steps may work as an efficient solution in the era of big data.

\section*{Acknowledgements}
The authors would like to express heartfelt gratitude to the anonymous reviewers and editor for their comments that improved the quality of this paper. This work was supported by and in part by the Natural Science Foundation of China under Grant Nos. 62071240 and 61802002, the Natural Science Foundation of Jiangsu Higher Education Institutions of China under Grant No. 19KJB520028, the Graduate Research and Innovation Projects of Jiangsu Province under Grant No. KYCX20\_0978, the Practice Innovation Training Program Projects for Jiangsu College Students under Grant No. 201910300140Y, and the Priority Academic Program Development of Jiangsu Higher Education Institutions (PAPD).


\end{document}